 \documentclass[preprint2]{aastex}
    
\shortauthors{Laurikainen et al.}

\begin{document}


\title{Bars, ovals and lenses in early-type disk galaxies: \\
    probes of galaxy evolution}


\author{E. Laurikainen\altaffilmark{1} and H. Salo\altaffilmark{1}}
\affil{Dept. of  Phys. Sciences/Astronomy Division, University of Oulu, FIN-90014, Finland}

\author{R. Buta\altaffilmark{2}}
\affil{Dept. of Physics and Astronomy, University of Alabama, Box 870324, Tuscaloosa, AL 35487}
%\email{aastex-help@aas.org}

\and

\author{J. H. Knapen\altaffilmark{3}}
\affil{Instituto de Astrof\'\i sica de Canarias, E-38200 La Laguna, Spain}


%%%\altaffiltext{1}{Visiting Astronomer, Cerro Tololo Inter-American Observatory.
%%%CTIO is operated by AURA, Inc.\ under contract to the National Science
%%%Foundation.}
%%%\altaffiltext{2}{Society of Fellows, Harvard University.}
%%%\altaffiltext{3}{present address: Center for Astrophysics,
%%%    60 Garden Street, Cambridge, MA 02138}
%%%\altaffiltext{4}{Visiting Programmer, Space Telescope Science Institute}
%%%\altaffiltext{5}{Patron, Alonso's Bar and Grill}


\begin{abstract}

The origin of S0 galaxies is discussed in the framework of early
mergers in a Cold Dark Matter cosmology, and in a scenario where S0s
are assumed to be former spirals stripped of gas. From an analysis
of 127 early-type disk galaxies (S0-Sa), we find a clear correlation
between the scale parameters of the bulge ($r_{\rm eff}$) and the disk
($h_{\rm R}$), a correlation which is difficult to explain if these
galaxies were formed in mergers of disk galaxies. However, the
stripping hypothesis, including quiescent star formation, is not
sufficient to explain the origin of S0s either, because it is not
compatible with our finding that S0s have a significantly smaller
fraction of bars (46$\pm$6 $\%$) than their assumed progenitors, S0/a
galaxies (93$\pm$5 $\%$) or spirals (64-69 $\%$). Our conclusion is
that even if a large majority of S0s were descendants of spiral
galaxies, bars and ovals must play an important role in their
evolution. The smaller fraction particularly of strong
bars in S0 galaxies is compensated by a larger fraction of
ovals/lenses (97$\pm$2 $\%$ compared to 82-83 $\%$ in spirals), many
of which might be weakened bars. We also found massive
disk-like bulges in nine of the S0 galaxies, bulges which might have formed at
an early gas-rich stage of galaxy evolution.

\end{abstract}





\section{Introduction}

The formation of S0 galaxies is generally discussed in the framework
of mergers of disk galaxies within a Lambda Cold Dark Matter
($\Lambda$CDM) cosmology. Alternatively, they can be viewed as former
spirals where star formation has ceased and gas lost by stripping
mechanisms. However, less attention has been paid to the
role of bars in the evolution of these galaxies. Bars are known to be
efficient drivers of gas towards the central regions of galaxies, and
in the presence of nuclear bars \citep{shlosman}
%(Shlosman, Frank $\&$ Begelman 1989) 
or nuclear spirals, a central starburst might occur.
Numerical simulations \citep{atha2,valpuesta2}
%(Athanassoula 2002, 2003) 
also predict that bars evolve due to angular momentum transfer between
the bar and a massive or centrally concentrated halo, leading to more
prominent bars. Indeed, there is observational evidence that bars in
early-type disk galaxies might be more evolved than bars in spiral galaxies. They
are longer \citep{elmegreen,laurikainen2007},
%(Elmegreen $\&$ Elmegreen 1985; Erwin 2005; Laurikainen et al. 2007),
more massive, and have more frequently double-peaked Fourier amplitude
profiles \citep{laurikainen2007}
%Laurikainen et al. 2007) 
and ansae-type morphologies \citep{laurikainen2007,valpuesta}.
%(Laurikainen et al. 2007; Martinez-Valpuesta et. al. 2007)
In the above simulation models these features can be interpreted as indices of
evolved bars.

Lenses are features commonly observed in S0 galaxies, but 
are sometimes also seen in early-type spirals. They are defined as
components with a shallow or constant surface brightness profile, and
a sharp outer edge. Lenses are a fundamental part of the original
classification of S0s \citep{sandage2}, although they were not recognized
initially with their own type symbol. Ovals are global deviations of the
disk from the axisymmetric shape. In distinction to bars they have
lower ellipticities and generally lack higher order Fourier terms
than $m$=2.% However, as the distinction between ovals and lenses is not 
%clear, in the following we call both types simply as ``lenses''.

We emphasize the importance of ovals and lenses for understanding bar-induced
galaxy evolution. \citet{kormendy} showed that 54$\%$ of barred galaxies
of types SB0-SBa have lenses. The actual frequency of lenses in
non-barred S0s was not determined at that time, although strong lenses
were known to exist in such galaxies \citep{kormendy2,sandage}.
Our recent studies (Laurikainen et al. 2005, 2007)
have confirmed that lenses are common in non-barred S0s.  It was first
suggested by \citet{kormendy} that lenses might be destroyed or
weakened bars.  If bars evolve in the Hubble sequence, and lenses are
indeed weakened bars, at some stage one would expect the fraction of
lenses to exceed the fraction of bars.  %However, lenses as a structure
%component of galaxies has received little attention: for example,
%their fraction has never been studied for any significant sample using
%deep near-IR images.  
It is also challenging to explain the origin of the multiple bars,
ovals and lenses seen in single S0s and for which no
explanation has yet been given in the current paradigm of galaxy
formation.
%by Petipas $\&$ Wilson (2004, ApJ, 603, 495) that SB0 galaxies often
%Data reduction in paper XX and the analysis of bulges in paper XX. 

In this Letter the origin of S0 galaxies is discussed, based on an
analysis of 127 early-type disk galaxies (Laurikainen et al. 2005,
2006): N(S0)=82, N(S0/a)=18, N(Sa-Sab)=27. Although by far most S0s
might be descendants of spiral galaxies, we present evidence that
bars, ovals and lenses have played an important role in their
structure formation and evolution.  Support for the stripped spiral
hypothesis is provided by a clear correlation between bulge and disk
scale parameters (Section 4).  The role of bars in structure formation
is evidenced by the large number of ovals+lenses (interpreted as
weakened bars) in S0s, and by the massive disk-like bulges found in
nine of the galaxies (Sections 3 and 5).  The studied galaxies are
part of a magnitude-limited Near-IR S0 galaxy Survey (NIRS0S), which
has the following selection criteria: $B_{\rm T}$ $\le$ 12.5,
inclination $\le$ 65$^\circ$, and Hubble type $-$3 $\le$ $T$ $\le$ 1
\citep{laurikainen2005,buta}.  More than half of the 184 NIRS0S
galaxies are currently analyzed, which form an adequate sample to
study the structural components of these galaxies.
%(Laurikainen et al. 2005; Buta et al. 2006).
%Except for Sa-galaxies, we have 16-26 galaxies in each Hubble type bin.

%$\ge$ $\le$
 \section{Two examples of oval/lens-dominated galaxies}

We show 2-dimensional decompositions of the surface brightness
distribution for two oval/lens-dominated galaxies, with the following
aims: (1) to show how to identify lenses in galaxies, (2) to stress
the importance of accounting for bright ovals/lenses while deriving
the parameters of the bulge, and, (3) to demonstrate the complexity of
some S0 galaxies that needs to be explained by galaxy evolutionary
models.  Our decompositions use a Sersic function for the bulge,
allowing for its flattening and deviation from elliptical isophotes,
an exponential function for the disk, and either a Sersic or Ferrers
function for the bars, ovals and lenses. In this study two types of
bulges are considered: (1) classical elliptical-like bulges (with
Sersic index $n$ near 4), and (2) disk-like pseudobulges (with smaller
Sersic index), formed mainly from the disk material via central star
formation.  However, the vertically thick boxy/peanut structures in
barred galaxies (often also called bulges) \citep{atha2} are
considered here as part of the bar. Such structures, and
nuclear bars and rings are not included to the flux of the bulge in
our decompositions.
%This definition of a bulge has implications
%for our decomposition approach, because nuclear bars and rings are counted
%as part of the disk.

The galaxies and their decompositions are shown in Figures 1 and 2: NGC 524 is
dominated by two almost circular lenses, whereas NGC 5365 has
two oval-shaped components, both ovals with an embedded bar.
%The distinction between an oval and a lens is not clear, but we assume
%that ovals are brighter and have larger vertical thicknesses than
%lenses.  
The lenses in NGC 524 are directly visible in the image, and showing
in the surface brightness profile as distinct exponential
sub-sections. The best fit is obtained by $n$=2.8 Sersic bulge and two
flat Ferrers functions (Ferrers index=1) for the lenses, implying a
fairly small bulge-to-total flux ratio $B/T$=0.28.  For NGC 5365 the
inner bar/oval system is fitted by a single Sersic function and the
outer one with a single Ferrers function, leading to a fairly
exponential bulge ($n$=2.0) and a small $B/T$=0.17. Counting all the
flux above the exponential disk as a bulge, 
would lead to a considerable overestimate of the bulge flux
($B/T$ $\sim$ 0.5 in two component fits), in accordance with Laurikainen et
al. (2005) who showed that by omitting the bar/oval in the decomposition the
$B/T$-flux ratio is overestimated, regardless of whether 1D or
2D-decompositions are used. It is challenging to explain how this kind of multiple
ovals/lenses form in S0 galaxies and how they are related to the evolution
of bars. In Section 6 they will be discussed in the context of cosmologically
motivated simulations by Heller et al. (2007). 
%Interestingly, the surface brightness profile of NGC 5365
%with its lens signatures resembles those obtained for the
%evolved stages of the first generation bars in the simulations by
%\citet{heller}, who studied the evolution of disks immersed in
%triaxial dark matter halos.
%Heller, Shlosman $\&$ Athanassoula (2007).

\section{Weak bars inside the lenses}

Although bars in S0 galaxies are on average fairly prominent, weak bars
are detected inside the ovals/lenses in eight of the galaxies: NGC 484, NGC %108
507, NGC 1161, NGC 1351, NGC 2768, NGC 2902, NGC 3998 and NGC 7377.
All these galaxies are classified as non-barred (de Vaucouleurs et
al. 1991; RC3) and no bar is directly visible in the $K_{\rm s}$-band
image. However, a weak, genuine bar is visible in the residual image
after subtracting the bulge model obtained from our 
decomposition (see Fig. 3 for NGC 3998).  Taking into account that
the evolution of bars and bulges might be coupled, it is interesting
to look closer at the properties of the bulges of these galaxies.

%\citep{laurikainen2008}. 
%(Laurikainen et al. 2008) 
We find fairly small Sersic indexes for the bulges in these nine
galaxies ($<n>$=2.5), similar to the typical values recently found for S0
galaxies in decompositions where a multi-component approach is used
(Laurikainen et al.  2005, 2007; Gadotti 2008).
%\citep{balcells1,balcells}.
%(Graham 2001; Balcells et al. 2003). 
However, their average bulge-to-total flux ratio ($<B/T>$ = 0.44) 
is higher than average for their Hubble type, which for
S0-S0/a galaxies $\sim$ 0.25-0.28 (Laurikainen et al. 2005, 
2007; Gadotti 2008).  In Section 6 we will discuss
that the weak bars, and probably also the massive, fairly
exponential bulges in these galaxies, might be a manifestation of
bar-induced secular evolution.

\section{Comparison of the scale parameters of the bulge and the disk}

The scale parameters of the bulge and the disk are sensitive to the
evolutionary processes of galaxies during their cosmic history, and
therefore offer an independent test of the importance of secular
processes in S0s. A correlation between the scale parameters is
expected in models where the bulges were formed in slow secular
processes (reviewed by Kormendy $\&$ Kennicutt 2004),
%(see Kormendy $\&$ Kennicutt 2004),
whereas no correlation is expected in models where the bulges were
formed either in a fast dissipative collapse \citep{eggen}, or
%(Eggen, Lynden-Bell $\&$ Sandage 1962)
by mergers of large \citep{toomre,springer}
%(Toomre $\&$ Toomre 1972; Springer $\&$ Hernquist 2005) 
or small galaxies \citep{abadi}.
%(Abadi et al. 2003).

The radial scale length of the disk, $h_{\rm R}$, and the absolute
magnitude, $M$, in the $K_{\rm s}$-band are calculated in the following manner:
\vskip 0.25cm
%$r_{eff,intr}$ [kpc] = $r_{eff,obs}$ * Dist

$h_{\rm R,intr}$ [kpc] = $h_{\rm R,obs}$ $\times$ D[kpc] / $\rm c_1$. 

$M_{\rm intr}$ = $m_{\rm obs}$ - $m_{\rm ext}$ - $\rm c_2$ - (5 $\times$  log D[Mpc]) - 25.0,

\vskip 0.25cm

\noindent 
where the internal dust correction for $h_{\rm R}$, $\rm c_1$ = [1.02 - 0.13 $\times$ log
  (cos $i$)], is from \citet{grahamworley}.
%Graham $\&$ Worley (2008).
For $M$, the internal dust correction is
$\rm c_2$ = [0.11 + 0.79 $\times$ (1-cos $i$)$^{2.77}$], taken from \citet{driver}.
%Driver et al. (2008). 
%For $r_{eff}$ no internal extinction correction is made.  
For galaxy distance D we use $H_{\rm o}$ = 75 km s$^{-1}$ Mpc$^{-1}$,
and, Galactic extinction, $m_{\rm ext}$, is from \citet{schlegel}.
%Schlegel et al. (1998). 
For the effective radius of the bulge, $r_{\rm eff}$, no internal dust
correction was made.  The subscripts $obs$ and $intr$ refer to
measured and intrinsic values, respectively.  We find a clear
correlation between $r_{\rm eff}$ and $h_{\rm R,intr}$, with the coefficient of
correlation of 0.66, at the significance level of 7E-14 (Fig. 4).
The correlation is independent of the applied corrections.  The
correlation is found to be the same for bright ($M$ $<$ -24.5) and
faint ($M$ $\ge$ -24.5) galaxies. This contradicts the recent result
by \citet{barway}
%Barway et al. (2007) 
who found an anti-correlation for bright S0s, and a positive
correlation for faint S0s. This different result is most probably due
to the deeper images and the more homogeneous database used in the present
study.  Notice also that the scale parameters in Barway et al.  were
derived using simple bulge-disk decompositions, whereas we use a
multi-component approach.  The correlation we find for S0s is similar
to that found previously for spiral galaxies
\citep{courteau,carollo},
%(Khosroshahi et al. 2000;
%Courteau et al. 1996; Carollo et al. 2007), 
and for 14 SB0 galaxies \citep{aguerri}.

\section{Frequency of bars, ovals and lenses}

%% In a manner similar to \objectname authors can provide links to dataset
%% hosted at participating data centers via the \dataset{} command.  The
%% second curly bracket argument is printed in the text while the first
%% parentheses argument serves as the valid data set identifier.  Large
%% lists of data set are best provided in a table (see Table 3 for an example).
%% Valid data set identifiers should be obtained from the data center that
%% is currently hosting the data.
%%
%% Note that AASTeX interprets everything between the curly braces in the 
%% macro as regular text, so any special characters, e.g. "#" or "_," must be 
%% preceded by a backslash. Otherwise, you will get a LaTeX error when you 
%% compile your manuscript.  Special characters do not 
%% need to be escaped in the optional, square-bracket argument.

%% In this section, we use  the \subsection command to set off
%% a subsection.  \footnote is used to insert a footnote to the text.

%% This section also includes several of the displayed math environments
%% mentioned in the Author Guide.

Finally, we compare the fractions of galaxies with bars and
ovals/lenses in different Hubble types. If S0s were simply stripped
spirals one would expect similar bar fractions in S0s and in their
spiral progenitors. In Table 1 we use RC3 family classes for
calculating bar fractions in different Hubble type bins: for S0-S0/a
galaxies the whole NIRS0S sample of 184 galaxies is used, whereas for
spirals we use the similarly sized Ohio State University Bright Spiral
Galaxy Sample (OSUBSGS, Eskridge et al. 2000). We find that S0s
(46$\pm$5 $\%$) have bars (SB+SAB) less frequently than S0/a galaxies
(77$\pm$9$\%$) or spirals (61-70 $\%$). The values for spirals are in
agreement with those found previously by other authors
\citep{knapen,eskridge,laurikainen2004,delmestre,marinova}. The bar
fractions for S0 and S0/a galaxies in the sub-sample of 127 NIRS0S galaxies
(38$\pm$5 and 76$\pm$10 $\%$, respectively) are nearly the same as for
the complete NIRS0S sample. In Table 1 the fractions of multiple bars are
calculated in respect of the total number of barred galaxies, while
all the other values are given in respect of the total number galaxies
within the Hubble type bin. The uncertainties are estimated from
$\Delta p = \sqrt{(1-p)p/N}$, where $p$ denotes the fraction in
question in a sample of $N$ systems.

We then use the $A_2$, the maximum m=2 Fourier density amplitude in the bar
region, normalized to m=0, to study three bar strength bins
within each Hubble type bin. For the OSUBSGS we use the values from
Laurikainen et al. (2004), whereas for S0 and S0/a galaxies they were
calculated in this study in a similar manner. As a lower limit for
the barred galaxies we use $A_2$=0.1.
%Notice that due to the
%lower limit of $A_2$=0.1 used for barred galaxies, not all weak bars discussed in
%Section 3 are included to the statistics.  
We confirm the above result that S0 galaxies have a smaller fraction
of bars than S0/a galaxies or spirals.  We find that: (1) Sc-Scd
spirals have the largest number of weak bars ($A_2$ = 0.1-0.3) and the
smallest number of strong bars (($A_2$ $>$ 0.6), and that the fraction
of strong bars increases towards the S0/a galaxies. This is in
agreement with the previous studies showing that the prominence of
bars increases towards the early-type disk galaxies. (2) Quite
interestingly, although the fraction of strong bars increases from
late-type spirals towards S0/a galaxies, it suddenly drops from
38$\pm$9 $\%$ to 10$\pm$3 $\%$ for S0s.

We also find that (3) S0 galaxies have a larger fraction of
ovals/lenses than S0/a galaxies galaxies (97$\pm$2 v.s 82$\pm9$ $\%$),
and that (4) S0/a galaxies have a larger fraction of multiple bars
than earlier or later type galaxies (see Laine et al. 2002; Erwin
$\&$ Sparke 2002). The fraction of ovals/lenses is found to be
the same for barred and for non-barred S0-S0/a galaxies (82$\pm4$ $\%$
v.s 86$\pm6$ $\%$, respectively). Although ovals and lenses might have
different light distributions, they are not distinct enough to be
considered separately in our statistics.

\section{Discussion and conclusions}

In the current paradigm of galaxy formation, $\Lambda$CDM, the
spheroidal components of galaxies were formed through mergers of disk
galaxies: dry mergers are suggested to lead to the formation of
elliptical galaxies, whereas mergers of gas poor with gas rich galaxies
lead to the formation of bulges in the disk-dominated galaxies
(Khockfar $\&$ Burkert 2003; Naab et al. 2006). Minor mergers are
actually more common in the Universe and they are suggested to form
even 55 $\%$ of the spheroid stars from accreted satellites (Abadi et
al. 2003). In $\Lambda$CDM the disks form after a major merger when
hot gas in the halo settles into the disk (Kauffmann et al. 1999;
Springer $\&$ Hernquist 2005). In this picture every dark matter halo
is expected to possess a substantial pressure supported classical
bulge with elliptical-like photometric properties (Steinmetz $\&$
Navarro 2002). Although many observations support this scenario it is
also faced with severe problems, for example the bulges, not only in
spiral galaxies, but even in S0-galaxies are fairly disk-like and have
smaller bulge-to-total mass flux ratios than predicted by cosmological
models.  {\it Our finding that the scale parameters of the bulge
  ($r_{\rm eff}$) and the disk ($h_{\rm R}$) are well correlated} for
S0s, provides an additional problem for $\Lambda$CDM: such a
correlation is difficult to explain if the formation of bulges and
disks in S0 galaxies were decoupled.

Alternatively, S0s might be descendants of spirals whose star
formation has faded after consuming the gas or losing it by some stripping
mechanism, like ram pressure stripping \citep{gunn},
%(Gunn $\&$ Gott 1972), 
halo stripping \citep{bekki},
%(Bekki et al. 2002), 
or galaxy harrasment \citep{moore}.
%(Moore et al. 1998). 
Recent evidence supporting this idea comes from the Tully-Fisher (TF)
relation and from the analysis of the properties of globular clusters in
galaxies. S0s lie below the spiral galaxies in the TF relation, having lower
luminosities \citep{bedregall}.
%(Bedregall et al. 2006), 
%and the globular clusters are a factor of 3 fainter than those in
%spiral galaxies
%(Aragon-Salamanca, Bedregal $\&$ Merrifield 2006; Barr et al. 2007).
%\citep{aragon,bedregall,barr}.  
This deviation is explained by luminosity evolution of spiral
galaxies: the transformation from spirals to S0s occurred at various
times in the past, and the galaxies have been passively fading ever
since. The globular cluster frequency (the number of globular clusters
per unit $V$-band luminosity) has been used as an independent estimate
of the degree to which the luminosity of S0s has faded relative to
that of its spiral progenitor \citep{aragon,barr}. This
estimate is based on the assumption that the frequency of globular
clusters is constant during the transformation process.
The fact that the bulges in S0-galaxies also have many characteristics
of disk-like structures, including their kinematic properties
(Cappellari et al. 2007), is consistent with this picture.  However,
if S0s were simply passively formed from S0/a spirals it would be
difficult to explain our finding that {\it the fraction of bars is
  considerably lower in S0s than in S0/a galaxies, which are expected
  to be their progenitors in the Hubble sequence. } Bars should be
fairly robust structures, evidenced by the fact that the bar fraction,
at least in massive luminous spirals, is maintained nearly constant
between the redshift range $z$= 0 - 0.84 \citep{sheth2}.
%(Sheth et al. 2008). 

Although the hypothesis of S0s as stripped spirals is a promising
idea, an important piece of information is still missing in this
picture.  Indeed, bars are expected to be efficient drivers of galaxy
evolution: the angular momentum transfer between gas and stars leads
to gas infall and subsequent star formation in the central regions,
which can add to the mass of the bulge \citep{friedli}.
%(Friedly $\&$ Benz 1993). 
If the angular momentum transfer occurs between the bar and the halo,
that leads to the evolution of the bar \citep{debattista,atha2}: a bar
first grows in mass and length, but if the bulge mass at the same time
increases due to the gas infall, that might lead to a subsequent
weakening of the bar.  Weakening of the bar is most efficient in
strong bars with flat-top surface density profiles \citep{atha3},
typical for early-type disk galaxies (Elmegreen $\&$ Elmegreen 1985).
In this study we have shown indirect observational evidence of such
evolution: {\it it is tempting to think that bar weakening due to
  increased central mass concentration is the explanation for the
  lower fraction of bars and the larger fraction of ovals/lenses in
  S0s.}  We also find that S0s galaxies have a deficiency particularly
of the strongest bars which fits to this picture.

{\it A manifestation of bar-induced secular evolution of galaxies is
  probably also our finding that nine galaxies in our sample have
  massive disk-like bulges ($<B/T>$=0.44, $<n>$=2.5), surrounded by
  weak bars and lenses}. In principle, an increase in bulge mass in
these galaxies could have occurred in a similar manner as discussed
above. Also, once the bulge mass had increased, the bar might have
started to weaken, leaving only a weak bar inside a lens. The lenses
surrounding the weak bars can be naturally explained as relics of the
evolution of the bar: many barred galaxies have lenses of the same
dimension as the bar, aligned with the bar major axis. A problem in
this scenario is that normal spiral galaxies don't have enough gas for
making such massive bulges by star formation (Kormendy $\&$ Kennicutt
2004), not at least without a significant accretion of extragalactic
gas to the disk.  It is still possible, however, that these galaxies
are associated with the formation and evolution of bars at higher
redshifts.  The role of bars in cosmological simulations has not yet
been studied much, but one such attempt has been made by
\citet{heller}.
%Heller, Shlosman $\&$ Athanassoula (2007). 
Their simulations, starting from initial values motivated by
cosmological simulations, include star formation, cooling and
feedback. In these simulations disk-like bulges form at early
phases of galaxy evolution during the gas-rich epoch in the history of
galaxies, being thus capable of accounting for the large
bulge-to-total flux ratios of the disky bulges found in some of the S0s.
The triaxial halos are the driving force to the formation of
primordial bars, which trigger nuclear bars.  
After 4-5 Gyr the primary bars are weakened to fat ovals and the nuclear
bars are decoupled. %It has been
%by \citet{atha3}
%Athanassoula (2008)
These processes might be a key for understanding the
multiple bar/oval/lens structures seen in many S0 galaxies.
%Namely, it has been shown by
%\citet{debattista} that disk-like bulges are sensitive to
%disk instability leading to formation of bars. The disks in S0s are
%not very cool, needed for the instability of the disk, but they still
%have some fraction of gas \citep{sheth}, which gas fraction must
%have been higher in the distant past.

Acknowledgements: E. Laurikainen and H. Salo acknowledge the Academy
of Finland for support, and R. Buta acknowledge the support of NSF Grant
AST-0507140.

%%%%%\appendix




\clearpage

\begin{figure}
\includegraphics[angle=0,scale=.60]{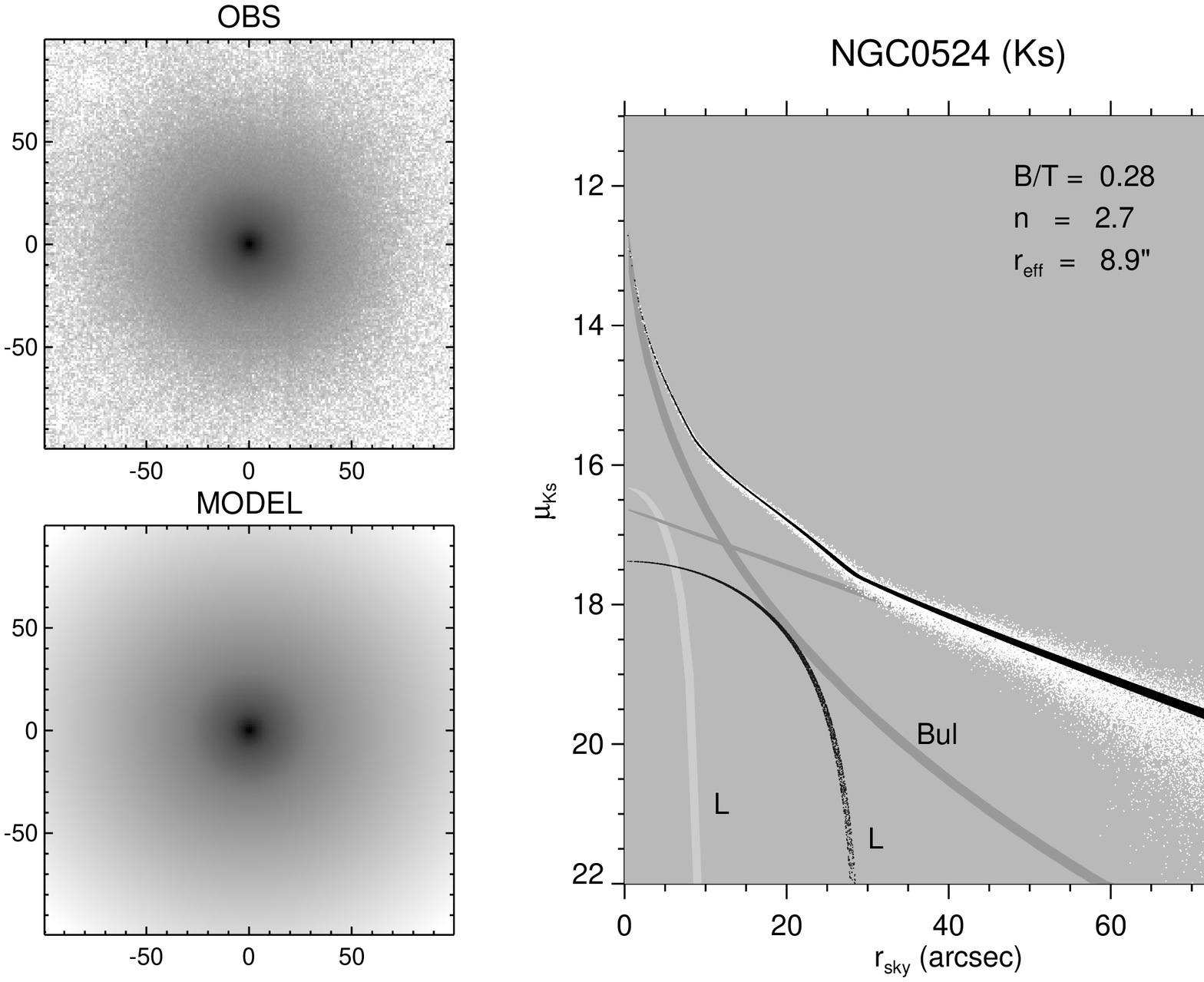}
\caption{Two-dimensional multi-component decomposition is shown for
  NGC 524.  Small white dots show the pixel values in the observed
  image and the other symbols show the model functions: dark grey
  lines are for the Sersic bulge (Bul) and the exponential disk, and
  small black dots show the final model. Lenses (L) are shown
  by large black dots and by shadowed light grey. The upper left panel
  shows the observed $K_{\rm s}$-band image and the lower panel is
  the fitted model image.
\label{fig1}
}
\end{figure}

\clearpage

\begin{figure}
\includegraphics[angle=0,scale=.60]{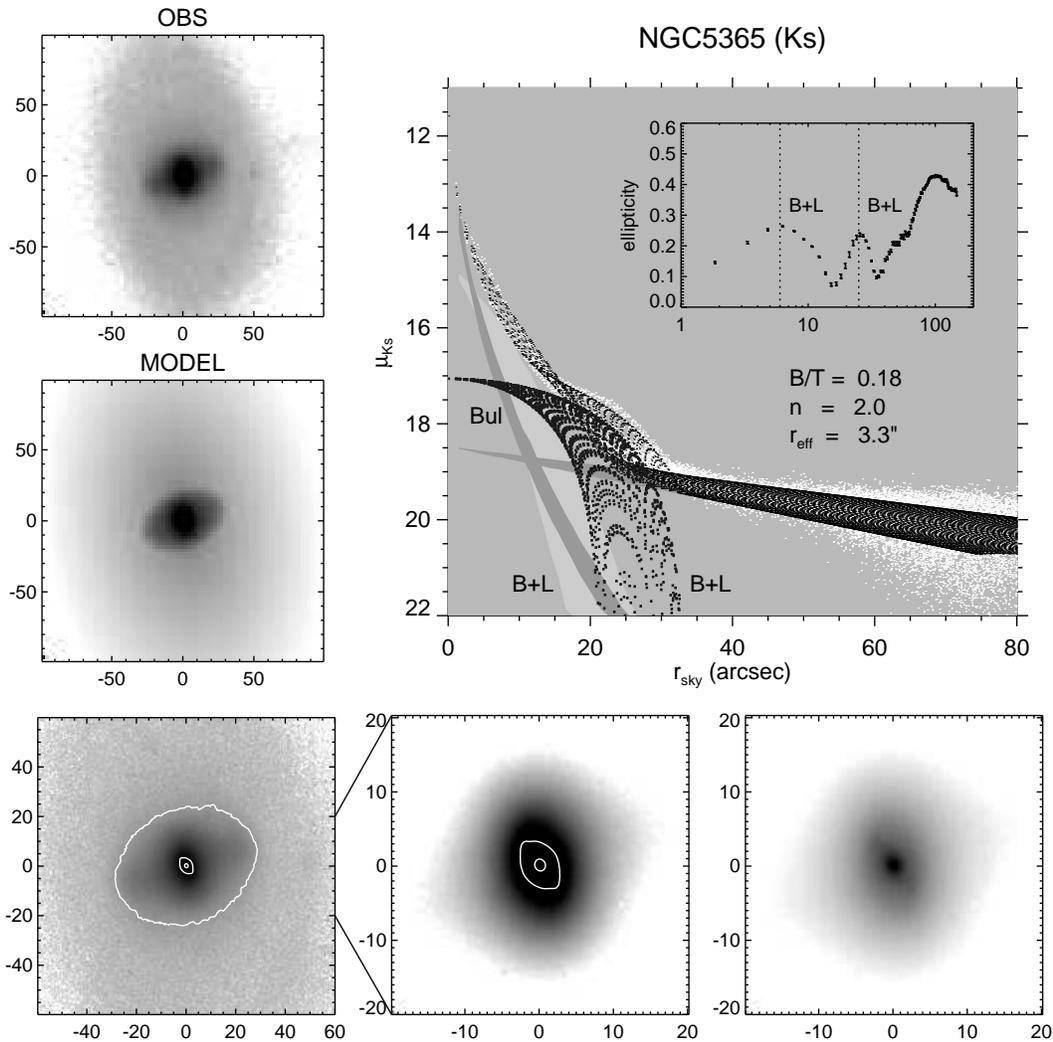}
\caption{ Decomposition for NGC 5365. The symbols are as in Fig.1.
  The galaxy has two bars embedded in ovals,
  both components being fitted by a single function (B+L).  The
  three lower panels show the observed image in different scales. {\it
    Left}: the scale is selected to show the primary bar, whereas the
  isophotes show the nuclear bar and the weak oval surrounding the
  primary bar. {\it Middle}: shows the bright inner oval, and an
  isophote indicating the nuclear bar.  {\it Right:} shows the nuclear
  bar. In the ellipticity profile the two bar+oval systems have nearly
  the same ellipticity.
%Notice also the outer ring in NGC 5365.
\label{fig2}}
\end{figure}

\clearpage

\begin{figure}
\includegraphics[angle=0,scale=.70]{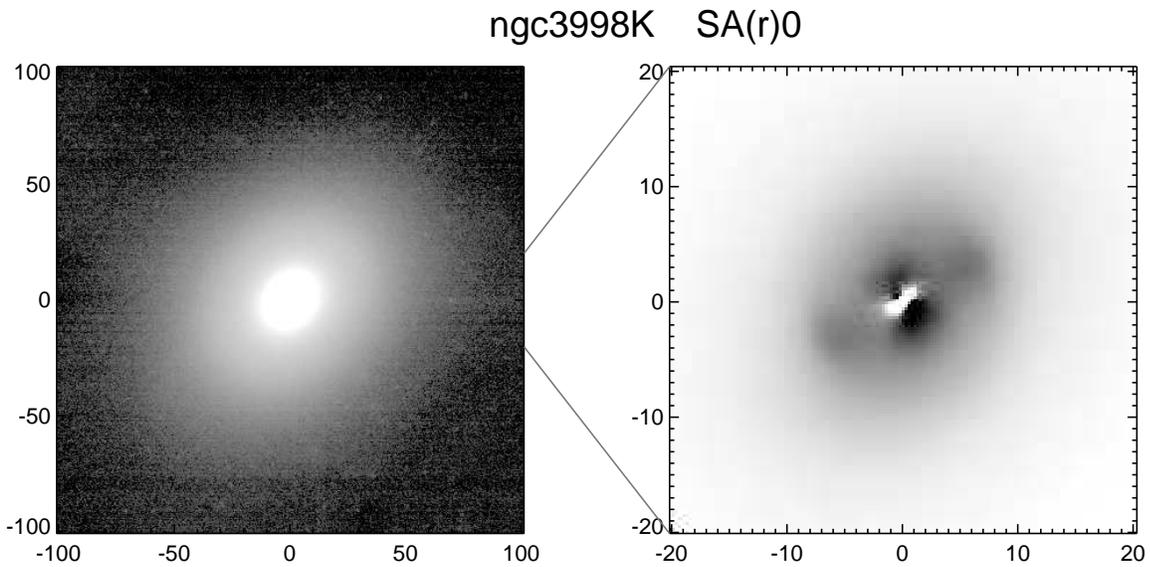}
\caption{ Left: the original $K_{\rm s}$-band image of NGC 3998; 
  Right: the residual image after subtracting the bulge
  model obtained from our 2D decomposition. The bright lens at $r$ $<$ 12''
  was fitted by a Ferrers function.  The scale is in
  arcseconds.
\label{fig3}}
\end{figure}

\clearpage

\begin{figure}
%\epsscale{.80}
%\plotone{FIG_1_letteri.eps}
%\plotone{Laurikainen_FIG_4.eps}
\includegraphics[angle=0,scale=.80]{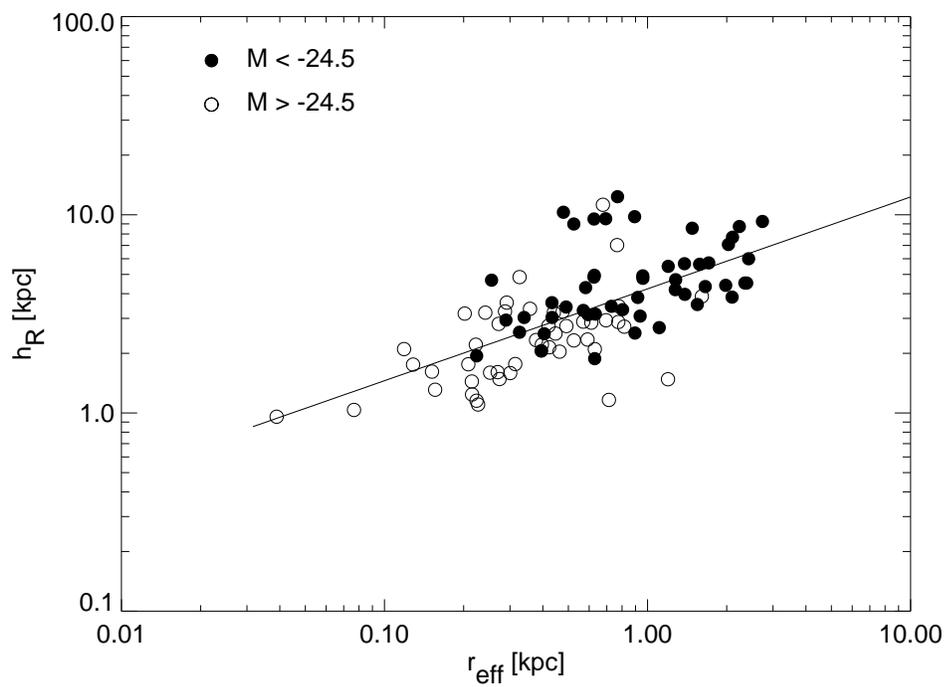}
\caption{The intrinsic scale length of the disk as a function of the
  effective radius of the bulge. Two
  $K_{\rm s}$-band magnitude bins are shown ($H_0$=75 km s$^{-1}$
  Mpc$^{-1}$ is assumed). The line shows the fit
  $\log h_{\rm R} = 0.62 + 0.46 \log R_{\rm eff}$.
    \label{fig4}}
\end{figure}

\clearpage

\begin{table*}
\small
\centering
  \caption{Bar fractions using RC3 family classes for the complete
    NIRS0S+OSUBGS samples. The A$_2$ fractions for S0 and S0/a
    galaxies, and the statistics for ovals/lenses and multi bars have
    been derived from the NIRS0S sub-sample of 127 galaxies.}
  %\begin{tabular}{@{}llllll@{}}
    \begin{tabular}{@{}lccccc@{}}
     &  &  &  &  & \\
  \hline
%       &  &  &  &  & \\
 bar-index &S0$^-$, S0$^0$, S0$^+$   & S0/a & Sa,Sab  & Sb,Sbc & Sc,Scd \\
 
      (1)  & (2) & (3) & (4) & (5) & (6) \\
 \hline
% & & & & & \\

B+AB (RC3)  &46$\pm$5 $\%$ &77$\pm$9 $\%$  &65$\pm$7 $\%$ &70$\pm$6 $\%$ &61$\pm$7 $\%$ \\
B (RC3)     &27$\pm$4 $\%$ &50$\pm$10 $\%$ &30$\pm$7 $\%$ &39$\pm$6 $\%$ &27$\pm$6 $\%$ \\
AB (RC3)    &19$\pm$4 $\%$ &27$\pm$9 $\%$  &35$\pm$7 $\%$ &31$\pm$6 $\%$ &34$\pm$7 $\%$ \\  
A (RC3)     &53$\pm$5 $\%$ &22$\pm$9 $\%$  &35$\pm$7 $\%$ &28$\pm$5 $\%$ &38$\pm$7 $\%$ \\

& & & & & \\

all bars (A$_2$$>$0.1)  &53$\pm$6 $\%$ &93$\pm$5 $\%$ &65$\pm$7 $\%$ &69$\pm$6 $\%$ &64$\pm$7 $\%$ \\
strong (A$_2$ $>$ 0.6)  &10$\pm3$ $\%$ &38$\pm$9 $\%$ &26$\pm$7 $\%$ &23$\pm5$ $\%$ &11$\pm$4 $\%$ \\
medium (A$_2$=0.31-0.6) &33$\pm$6 $\%$ &44$\pm$9 $\%$ &35$\pm$7 $\%$ &34$\pm$6 $\%$ &26$\pm$6 $\%$\\
weak (A$_2$=0.1-0.3)    &9$\pm$4 $\%$ &10$\pm$5 $\%$ &5$\pm$4 $\%$ &11$\pm$4 $\%$ &28$\pm$6 $\%$ \\

 & & & & & \\
ovals/lenses  &97$\pm$2 $\%$ &82$\pm$9 $\%$ &83$\pm$7 $\%$ & & \\
multiple bars  &21$\pm$6 $\%$ &40$\pm$12 $\%$ &26$\pm$8 $\%$ & &\\
(among barred)& & & & & \\
%& & & & & \\
 \hline

\end{tabular}
\end{table*}

%**************************
%A (RC3)        &53+5 (55/102) &22+9 (5/22)  &35+7 (15/43) &28+5 (18/64) &38+7 (18/47) \\
%AB (RC3)       &19+4 (19/102) &27+9 (6/22)  &35+7 (15/43) &31+6 (20/64) &34+7 (16/47) \\
%B (RC3)        &27+4 (28/102) &50+10(11/22) &30+7 (13/43) &39+6 (25/64) &27+6 (13/47) \\
%B+AB (RC3)     &46+5 (47/102) &77+9 (17/22) &65+7 (28/43) &70+6 (45/64) &61+7 (29/47) \\
%& & & & & \\
%A$_2$=0.1-0.3  &9+4 (6/64)    &10+5 (3/29)  &5+4 (2/43)   &11+4 (7/64)  &28+6 (13/47) \\
%A$_2$=0.31-0.6 &33+6 (21/64)  &44+9 (13/29) &35+7 (15/43) &34+6 (22/64) &26+6 (12/47) \\
%A$_2$ $>$ 0.6  &10+3 (7/64)   &38+9 (11/29) &26+7 (11/43) &23+5 (15/64) &11+4 (5/47) \\
%A$_2$=all bars &53+6 (34/64)  &93+5 (27/29) &65+7 (28/43) &69+6 (44/64) &64+7 (30/47) \\
% & & & & & \\
%ovals/lenses   &97+2 (64/66)  &82+9 (14/17)  &83+7 (20/24) & & \\
%double bars    &21+6 (10/48)  &40+12 (6/15)  &26+8 (7/27)  & &\\
%(among barred)& & & & & \\

%--------------------------------------------------------------
%Koko NIRS0S:
%
%        N(SA)            N(SAB)        N(SB)         N(SAB+SB)
%
%Sa      8/31(25+8%)   12/31 (39+9)   11/31 (35+8)   23/31 (74+8)  
%S0/a    5/22 (22+9%)  6/22 (27+9)    11/22 (50+10)  17/22 (77+9)    
%S0+     14/31 (45+9)  8/31 (26+8)    9/31 (29+8)    17/31 (54+9)
%S0o     21/42 (59+7)  8/42 (19+6)    13/42 (31+7)   21/42 (50+8)
%S0-     20/29 (68+8)  3/29 (10+6)    6/29 (21+7)    9/29 (31+8)

%*************************
\end{document}